The 7th International Conference on Emerging Data and Industry (EDI40)
April 23-25, 2024, Hasselt, Belgium

# Predicting the usability of mobile applications using AI tools: the rise of large user interface models, opportunities, and challenges

Abdallah Namoun[a,*], Ahmed Alrehaili[b], Zaib Un Nisa[c], Hani Almoamari[a], Ali Tufail[d]

*[a]Faculty of Computer and Information Systems, Islamic University of Madinah, Madinah 23, Saudi Arabia*
*[b]Department of Informatics, University of Sussex, Falmer, Brighton BN1 9QT, United Kingdom*
*[c]Department of Computer Science and Information Technology, Superior University, Lahore, Pakistan*
*[d]School of Digital Science, Universiti Brunei Darussalam, Tungku Link, Gadong BE1410, Brunei*

**Abstract**

This article proposes the so-called large user interface models (LUIMs) to enable the generation of user interfaces and prediction of usability using artificial intelligence in the context of mobile applications. To this end, we synergized an integrated framework for the effective testing of the usability of mobile applications following a selective review of the most influential models of mobile usability testing. Next, we identified and analysed 13 recent AI tools that generate user interfaces for mobile apps, and systematically tested these tools to identify their AI capabilities. Our striking findings demonstrate that current generative UI tools fail to address mobile usability attributes, such as efficiency, learnability, effectiveness, satisfaction, and memorability. Our large UI models' architecture proposes to leverage the capabilities of large language models, large vision models, and large code models to overcome the challenges of AI-driven UI/UX design and front-end implementations. This fascinating UI eco-system must be augmented with sufficient UI data and multi-sensory input regarding user behaviour to train the models. We anticipate LUIMs to create ample opportunities, like expedited frontend software development, enhanced personalised user experience, and wider accessibility of smart technologies. However, the research challenges hindering the UI generation and usability prediction of mobile apps include the seamless integration of complex generative AI models, semantic understanding of non-uniform visual designs, scarcity of UX datasets, and modelling of realistic user interactions.

* Abdallah Namoun. Tel.: +966 598009417.
*E-mail address:* a.namoun@iu.edu.sa

## 1. Introduction and motivation

The average person spends approximately five hours daily on their mobile phones [3]. In 2022, 255 billion new app downloads were registered, and a whopping 167 billion USD was spent on app stores, a drastic increase from 230 billion app downloads in 2021. Interestingly, artificial intelligence is projected to increase mobile app downloads by 10% in 2024. To continue fueling their revenues in a highly competitive and volatile market, mobile app companies need to dedicate significant efforts to the design of user-friendly interfaces and the usability of their applications.

Usability testing of mobile applications is inherently a complex and expensive process [1], yet rewarding in elaborating user requirements, identifying usability issues, and improving the quality of user experience [2]. Mobile usability testing encompasses several intertwined and laborious phases, including planning and designing the evaluation sessions, recruiting the intended users, conducting the testing sessions, and analyzing testing data to extract actionable insights [1]. However, the explosion and prevalence of AI tools in different domains could be exploited to automate phases of the frontend development and usability evaluation of mobile applications and devices efficiently.

Machine learning models can improve the mobile user interface (UI) and augment the user experience (UX) considerably through the personalization of the app's look and feel based on peoples' preferences and past actions, improvement of visual designs and graphics, and recommendation of user-specific content. Moreover, generative AI and AI chatbots, such as ChatGPT, Bing, Bard, and Google Gemini, are now integrated within diverse mobile apps to optimize the content and interactions. Such generative AI functionalities are predicted to disrupt the mobile apps market by a surge of 40% yearly [4]. Vivid examples include the growth of AI chatbots by a factor of 72 and increase of generative AI apps downloads by a factor of 9 in 2023. However, AI models will need to be provided with sufficient data regarding the user behavior and UI design patterns to produce the best UI designs.

Despite the potential advances AI can bring to UX design, it remains unclear how current AI tools are used to generate new UIs and predict the usability and quality of mobile applications. This research is motivated by four intriguing questions:

- Research question one: What are the main attributes (or dimensions) that should be considered when predicting the usability of mobile applications?
- Research question two: What are the features and weaknesses of current AI UI tools dedicated for the generation of mobile applications?
- Research question three: What are the essential components of AI UI generators?
- Research question four: What are the major challenges hindering the application of AI models in the generation of usable mobile interfaces?

## 2. Related works

### *2.1. Usability of mobile applications*

The software industry places a strong emphasis on the usability of mobile applications. A mobile application enhances peoples' life only if it is easy to use and meets their needs. User acceptance of a system is significantly impacted by its usability, which is a vital facet of non-functional software qualities. According to the International Organization for Standardization (ISO), usability is defined as "*the extent to which a product can be used by specified users to achieve specified goals with effectiveness, efficiency, and satisfaction in a specified context of use*". Mobile application usability pertains to the use and circumstances of mobile devices, with the aim of enabling mobile users to complete a range of activities successfully [5]. Currently, there are 6.648 billion mobile users globally; 83.72% of whom use a smartphone [6]. The mobile app industry is a prosperous sector, catering to the needs of end users and facilitating the creation of diverse business models. The concept of mobile app usability arose concurrently with the development of mobile applications, aiming to assist both the developers and mobile users [7].

### 2.2. Automated mobile app usability testing

A range of testing methodologies is used to ensure the quality of software, including mobile applications [17]. Among those, automated usability testing has attracted a growing interest from software development companies. The challenge is in delivering exceptional user experiences within the constraints of a small screen and a multitude of functions. Automated models offer a streamlined and time-saving approach for developers, potentially enhancing efficiency in usability testing. Automated usability testing tools aim to reduce costs, time, and human resources by replacing human experts with computerized evaluations. Consumers are requested to test mobile applications and simultaneously use automated technologies for conducting usability tests. Usability concerns are identified via the analysis of user interactions with mobile apps, focusing on indicative patterns. Such tools rely on interaction design concepts to detect and resolve usability flaws [18]. Table 1 lists some popular automated tools for testing the usability of mobile applications.

Table 1: Example automated usability evaluation tools

| Automated Mobile Usability Tool | Goal and Application |
|---|---|
| User Testing | User Testing facilitates the creation of test scenarios and identifies the target consumers. Upon user engagement with the mobile app, this tool diligently logs their actions and collects their feedback. |
| Lookback | Lookback enables the conduction of remote user testing and usability analysis. Developers can view and engage with users as they navigate through the mobile application. |
| PlaybookUX | PlaybookUX makes it easy to conduct mobile app remote user testing. One can find video recordings, usability tests, and user-generated insights. |
| UserZoom | UserZoom provides usability testing and research solutions for mobile apps that includes testing, surveys, and analytics. |
| Userlytics | Userlytics helps with performing mobile and tablet testing. Both video footage and user comments are accessible. |
| CrazyEgg | Crazy Egg provides heatmaps and user behavior analytics and supports mobile app A/B and usability testing. |

However, automated usability testing fails to capture the nuanced emotional reactions that may be discerned by a human expert. Furthermore, it is challenging to replicate real-life situations. The semantics underlying user experience cannot be achieved due to the disparity between the analysis conducted by a human expert and the analysis emulated by automated techniques. Automated testing relies on scripted scenarios, and if these scenarios do not align with real-life situations, user behaviors will not be accurately reflected. The level of responsiveness in real-world settings differs from that in scripted scenarios. Users may experience interaction difficulties due to the absence of a responsive interface across different mobile applications. Moreover, users sometimes find it frustrating when they encounter new (unfamiliar) user interfaces in mobile learning apps [19]. Artificial intelligence-based usability solutions may be leveraged to adjust the UX and overcome user concerns and frustrations.

### 2.3. UI generation and usability prediction using artificial intelligence

Artificial intelligence (AI), machine learning (ML), and deep learning (DL) models are applied to a multitude of domains. The usability of mobile applications is no exception. These models can understand, learn, predict, and react like humans in similar situations and possess the qualities of machines, i.e., high processing speed and better accuracy. ML and DL algorithms may identify areas where the mobile applications fall short of user expectations and recommend fixes. These algorithms learn from the huge data that is gathered by user responses and feedback and learn how to improve the user experience, effectiveness, efficiency of mobile applications, covering nearly all the dimensions of usability [20]. Several studies show the use of AI based usability in their respective domain. In one study machine learning is used to find the suitability index for residential apartments [16], which is a usability dimension in the case of mobile apps showing prices of residential apartments. Soui et al. divided mobile designs into good and bad user interface designs and used extensive deep learning and machine learning algorithms to evaluate the quality of mobile apps [2]. Weimer et al. used a Convolutional Neural Network to identify defects in the industrial dataset of images [21]. In the same fashion, DL algorithms can be used to find the defects from the User Center Design (UCD) and can be very helpful in predicting usability. Machine learning is also used for customizing interfaces [22].

Generative AI is used to enhance the user experience and design of mobile applications [23]. Oztekin et al. [24] emphasizes the implementation of a continuous cycle until the user's requirements are met and satisfaction is achieved. For this reason, a cyclic process to achieve at a certain level is a specialty of generative AI. Generative AI has immense promise in elevating the usability of mobile apps to a higher degree via the utilization of large language models and vision models. Upon getting images of mobile apps, generative AI, particularly multimodal LLMs (Language and Vision Models), such as ChatGPT 4, Microsoft Copilot, and Gemini, can be used to identify usability problems and provide creative solutions.

Recently, LVMs have gained significant attention within the research community. The popularity of LVMs is boundless owing to its ability to express meaning that words alone cannot adequately describe, as exemplified by the well-known phrase "A picture is worth a thousand words". Nevertheless, the complete potential of LVM has not yet been fully explored. Significant progress has been made in the field of LVM, with several proposals being put forth. A research study introduces a vision language model (VLM) called "CogAgent" with 18 billion parameters, specifically designed for graphical user interface and navigation tasks. This model takes screenshots as input and provides users with clear instructions on how to navigate the graphical user interface and complete their tasks effortlessly [25]. This model facilitates the achievement of optimal user experience by offering user assistance and contributing to the attainment of usability objectives. In another study, the authors used LLMs for generating UI layouts [26]. A Generative Adversarial Network (GAN) has been used successfully to predict the next UI element in each layout [27] and for learning UI design patterns [28]. Genetic algorithms were used to design generative User Interface [29], showing that multiple diverse approaches are being used to enhance usability by providing the best user experience.

## 3. Results and discussion

### *3.1. Usability models and attributes for mobile applications*

Mobile applications have significantly improved people's lives by providing convenient access to a wide range of online services in diverse domains, like healthcare, education, social networking, and gaming. However, it is crucial to ensure the usability of these mobile applications to achieve user acceptance and adoption [8]. We performed a targeted analysis of the most prominent mobile usability models in the literature to answer our first research questions (see Table 2). Across 9 prominent survey studies, we identified 42 attributes for measuring the usability of mobile apps. The analysis showed that efficiency, learnability, effectiveness, satisfaction, and memorability are the most important elements of usability of mobile applications.

Table 2: Usability dimensions and attributes for mobile applications used in 9 prominent survey studies. In grey our selected mobile usability attributes.

| Usability Dimension/ Attribute for Mobile Applications | Harrison et al. [8] **Cited: 934** | Weichbroth P [9] **Cited: 145** | Alturki R et al. [10] **Cited: 37** | Baharuddin R et al. [11] **Cited: 245** | Hoehle H et al. [12] **Cited: 125** | Zhang D et al. [13] **Cited: 192** | Ismail N et al. [14] **Cited: 33** | Balagtas-Fernandez F [15] **Cited: 95** | Hoehle H et al. [16] **Cited: 577** | Total (out of 9) |
|---|---|---|---|---|---|---|---|---|---|---|
| Efficiency | ✓ | ✓ | ✓ | ✓ | ✓ | ✓ | ✓ | ✓ | ✓ | 9 |
| Learnability/ Learning Performance | ✓ | ✓ | ✓ | ✓ | ✓ | ✓ | ✓ | ✓ | ✓ | 9 |
| Effectiveness | ✓ | ✓ | ✓ | ✓ | ✓ | ✓ | | ✓ | ✓ | 8 |
| Satisfaction | ✓ | ✓ | ✓ | ✓ | ✓ | ✓ | | ✓ | ✓ | 8 |
| Memorability | ✓ | ✓ | ✓ | ✓ | ✓ | ✓ | | ✓ | ✓ | 8 |
| Errors | ✓ | ✓ | ✓ | | ✓ | ✓ | | ✓ | ✓ | 7 |
| Simplicity | | ✓ | ✓ | ✓ | ✓ | ✓ | | ✓ | | 6 |
| Ease of use | | ✓ | ✓ | | ✓ | | ✓ | | ✓ | 5 |
| Aesthetics | | ✓ | ✓ | ✓ | ✓ | | | | ✓ | 5 |
| Comprehensibility (Readability) | | ✓ | | ✓ | | ✓ | | ✓ | ✓ | 5 |
| Accuracy | | ✓ | ✓ | | ✓ | | | | ✓ | 4 |
| Operability | | ✓ | ✓ | ✓ | ✓ | | | | | 4 |
| Usefulness | | ✓ | ✓ | | ✓ | | | | ✓ | 4 |
| Attractiveness | | ✓ | ✓ | ✓ | ✓ | | | | | 4 |
| Consistency | | ✓ | | ✓ | ✓ | | | | ✓ | 4 |
| Cognitive load | ✓ | ✓ | ✓ | | ✓ | | | | | 4 |
| Enjoyment | | | ✓ | | ✓ | | ✓ | | | 3 |
| Understandability | | ✓ | ✓ | ✓ | | | | | | 3 |
| Visibility | | | ✓ | | ✓ | | ✓ | | | 3 |
| Interaction | | ✓ | ✓ | | | | | | | 2 |
| Accessibility | | ✓ | | ✓ | | | | | | 2 |
| Adaptability | | ✓ | | ✓ | | | | | | 2 |
| Flexibility | | | | ✓ | ✓ | | | | | 2 |
| Navigation | | ✓ | ✓ | | | | | | | 2 |
| Address user distraction | | | ✓ | | | | | | ✓ | 2 |
| User mobility | | | ✓ | | | | | | ✓ | 2 |
| Visual design | | | ✓ | | | | | | ✓ | 2 |
| Search Time | | | ✓ | | | | | | ✓ | 2 |
| Familiarity | | | | | ✓ | | | | ✓ | 2 |
| Intuitiveness | | | ✓ | ✓ | | | | | | 2 |
| Feedback | | | ✓ | | ✓ | | | | | 2 |
| Predictability | | | | | ✓ | | | | ✓ | 2 |
| Novelty | | | | | ✓ | | | | | 1 |
| Acceptability | | | | ✓ | | | | | | 1 |
| Responsiveness | | | | | ✓ | | | | | 1 |
| Recoverability | | | | | ✓ | | | | | 1 |
| Customizability | | | | | ✓ | | | | | 1 |
| User Control | | | | | ✓ | | | | | 1 |
| Task completion time | | | ✓ | | | | | | | 1 |
| Weight of device | | | ✓ | | | | | | | 1 |
| Convenience | | | ✓ | | | | | | | 1 |
| Reliability | | | | ✓ | | | | | | 1 |
| **Total Attributes** | **7** | **21** | **28** | **18** | **26** | **08** | **05** | **08** | **18** | |

In the current trend of AI-powered advancements, usability evaluation techniques have continued to grow beyond conventional methods. It has reached unprecedented levels of development by providing endless support and abilities such as AI-driven recommendations, AI-generated reports, AI-powered analysis, and AI-powered assistants

- etc. Our study recognizes the increasing influence of AI-powered tools, so we conducted a thorough search for AI-powered tools to evaluate the usability of mobile applications as presented in the next section.

### *3.2. Analysis of AI-driven UI design tools*

We carefully selected and evaluated 13 AI-driven software tools. To minimize bias, our choice of AI-powered tools followed a systematic and objective approach. This process consisted of three stages: (1) conducting a comprehensive search (Google and Microsoft Bing) for the most effective and prominent AI tools according to online reviews; (2) examining related studies in the field to assess the reputation of each AI tool; (3) calculating the frequency of each tool's mention in both areas. The search parameters were carefully defined to capture potential tools, with keywords such as "mobile application AI-powered usability evaluation", "AI-powered usability tools for mobile apps," and "AI and machine learning tools for mobile applications".

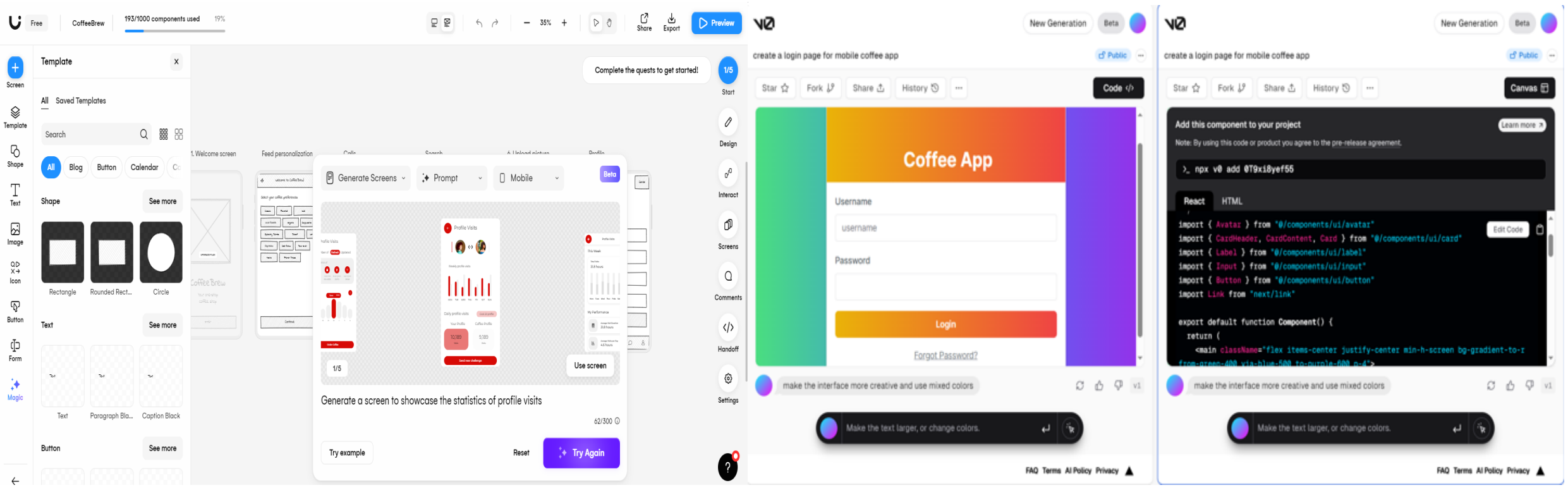

Fig. 1. Uizard tool creating a sequence of screens by text prompts (left) and V0 tool creating a design and its react code and html code (right).

Following these stages, we identified the top 13 AI-powered tools that met our criteria for an in-depth evaluation of their capabilities in testing the usability of mobile apps. Table 3 (Part I and II) summarizes the findings of the comparison across 19 criteria.

Table 3. A comparative analysis of 6 AI-powered UI generation tools – Part I

| AI tool name | Uizard | Smartlook | Maze | Visily | Optimizely | Lockback |
|---|---|---|---|---|---|---|
| Best for | Rapid prototyping and designing | Session Recordings, and Heatmaps. | Prototype evaluation and AI-powered interview summaries. | Mobile app usability testing and optimization | A/B testing and multivariate testing | One to one Interviews and live team collaboration |
| AI powered capabilities | Y | Partially | Partially | Y | Partially | Partially |
| AI Features | AI-driven designing capabilities of digital products, mobile apps, and website mockups. | AI-powered analysis of session recording. | AI-powered analysis of interviews transcription and summarizing. | AI-powered designing of software wireframes. | AI-powered personalized experiences and scenarios to different types of users. | AI-powered assistant in transcribing and summarizing of sessions |
| Free / Commercial / Research | Free trial /Commercial | Freemium | Freemium | Free | Commercial | Commercial |
| Does it require actual users? | N | Y | Y | N | Y | Y |
| Input - Text prompts | Y | N | N | N | N | N |
| Input - Hand drawn screens | Y | N | Y | Y | N | N |
| Input – Mockups | Y | N | Y | Y | N | N |
| Input – Code | N | N | N | N | N | N |
| Output - Low fidelity screens | Y | N | Y | Y | NA | NA |
| Output - High fidelity screens | Y | N | Y | Y | NA | NA |
| Output - Mobile websites | Y | Y | N | Y | Y | Y |
| Output - Hybrid apps | Y | Y | N | Y | Y | Y |
| Output - Native apps | Y | Y | N | N | Y | Y |
| Output – Usability Issues | Y | Y | Y | N | Y | Y |
| Mobile Apps – Smart Phones | Y | Y | Y | Y | N | Y |
| Mobile Apps – Tablets | Y | Y | Y | Y | N | Y |
| Does it support - Generative AI Functionality (text, image, video, audio etc.) | Y | N | Y | Y | Y | Y |
| Does it support - Analysis of User Data (Videos of interaction, Voice recordings of comments, Clicks) | N | Y | Y | N | Y | Y |

Table 3. A comparative analysis of 7 AI-powered UI generation tools – Part II

| AI tool name | UserTesting | Survcate | Global App Testing | Dynatrace | Galileo | Geniusui | v0 |
|---|---|---|---|---|---|---|---|
| Best for | Intelligent automation testing and AI driven analysis | In app customer feedback | Crowd testing | Monitoring and performance tracking | Rapid high-fidelity UI designs | Rapid UI designs | Rapid prototyping and designing |
| AI powered capabilities | Partially | Partially | Partially | Partially | Y | Y | Y |
| AI Features | AI and machine learning to analyze user behavior, emotions, and opinions. | AI-powered analysis of user feedback. | AI-powered detecting of visual differences and anomalies in apps or websites. | AI-powered to resolve performance issues in applications | AI-powered design of UI | AI-powered design of UI | Design digital products, mobile apps, and website mockups. |
| Free / Commercial / Research | Commercial /Research | Commercial | Commercial | Commercial | Waitlist/ Temporarily stopped | Waitlist/ Temporarily stopped | Free trial/Commercial |
| Does it require actual users? | Y | Y | Y | Y | N | N | N |
| Input - Text prompts | N | Y | N | N | Y | Y | Y |
| Input - Hand drawn screens | N | N | N | N | N | N | Y |
| Input – Mockups | N | N | N | N | N | N | Y |
| Input – Code | N | N | N | N | N | N | N |
| Output - Low fidelity screens | NA | NA | NA | NA | Y | Y | Y |
| Output - High fidelity screens | NA | NA | NA | NA | Y | Y | Y |
| Output - Mobile websites | Y | Y | Y | Y | N | N | Y |
| Output – Hybrid apps | Y | Y | Y | Y | N | N | Y |
| Output - Native apps | Y | Y | Y | Y | N | N | N |
| Output – Usability Issues | Y | Y | Y | Y | N | N | N |
| Mobile Apps – Smart Phones | Y | N | Y | N | Y | Y | Y |
| Mobile Apps – Tablets | Y | N | Y | N | Y | Y | Y |
| Does it support - Generative AI Functionality (text, image, video, audio etc.) | N | N | N | N | Y | Y | Y |
| Does it support - Analysis of User Data (Videos of interaction, Voice recordings of comments, Clicks) | Y | Y | Y | Y | N | N | N |

The results demonstrate that each AI-based development tool has its own unique characteristics in terms of predicting mobile usability. There are two methods for measuring mobile apps' usability: through actual user monitoring or with predefined rules and guidelines. Tools like "UserTesting" and "Surrogate" depend on actual user data for analysis, deploying user activity monitoring to evaluate the usability of apps. Conversely, platforms such as "Uizard", "v0" and " Visily" function without direct user interaction, depending on predefined rules or automated assessments.

Regarding tools' accessibility, "Uizard" offers a free trial, and " Geniusui " is available both as a free trial and a commercial product. While most tools operate on a commercial basis, some offer more accessible options like freemium models and free trials. However, "Galileo" and "Geniusui" are currently unavailable, suggesting their reliance on other models in high demand.

In term of supported input, tools like "Uizard," "Galileo," "Geniusui," and "v0" support text prompts for rapid visualization. Others, including "v0", "Uizard", "Maze", and "Visily", assist novice designers by supporting hand-drawn screens and mockups. These features enable designers to transform their initial ideas into designs in a dynamic way. Conversely, "Smartlook" and "Optimizely" do not support mockups, focusing on usability evaluation during analytics and testing stage rather than the initial designing stage. Notably, none of the tools listed explicitly accept code as an input, which suggests the complexity of integrating this feature.

In terms of the supported output, tools like "Uizard," "Maze," and " Geniusui" provide both low and high-fidelity designs, allowing designers to progress from basic outlines to detailed designs. Many tools, including "Uizard" "Smartlook", "Maze", and "Visily" support mobile website outputs, which is critical for our study in investigating the usability in mobile apps and mobile websites. Furthermore, "Uizard" stands out due to its versatility, offering output designs for mobile websites, hybrid apps, and native apps across devices, unlike "Lookback" and "Optimizely" which lack this diversity.

In summary, the study explores a range of AI-powered tools. Some of these tools are data-driven platforms like "UserTesting" and "Smartlook", which analysis actual user feedback to enhance usability. Other tools support autonomous design like "Uizard" and " Visily", which utilizes AI to accelerate the design process. These various tools, spanning from freemium to commercial models, provides several options for designers. Their capability to handle diverse inputs and produce a wide range of output types, illustrates the powerful capabilities of such AI - powered tools in meeting the evolving the usability mobile applications.

In attempting to answer the second part of the second question (i.e., What are the features and (weaknesses) of the current AI tools for mobile application generation?), the results show that although AI-powered mobile application tools provide general insights on how to enhance usability, there are some weaknesses that need to be addressed. Below, we present a set of weaknesses that need to be resolved to enhance the efficiency of AI-powered mobile application evaluation tools.

- Weakness One (Training Large UI Models): The tools listed do not seem to offer capabilities for training machine learning models, which could be a significant limitation for organizations looking to leverage AI for continuous improvement of their UI/UX.
- Weakness Two (UI/UX Design Practices and Patterns): None of the AI tools indicate if they offer guidance or adherence to established design patterns and best practices, which is a key aspect of design and development according to well-known usability standards.
- Weakness Three (Quality Testing and Validation): While the tools provide various testing capabilities, they do not explicitly mention quality assurance or validation services, which are essential for ensuring that the UI meets certain quality standards before deployment.
- Weakness Four (Source Code Integration): None of the tools explicitly accept code as an input, which suggests a potential gap in accommodating designers who may want to integrate existing code directly into the design and test its output.
- Weakness Five (Output Limitations): A common limitation across several tools is the absence of output for native mobile apps, which is a critical requirement for creating optimized applications for specific platforms like iOS or Android.

In conclusion, while the tools offer a range of AI-powered features for evaluating the usability of mobile applications, they appear to lack direct capabilities in source code integration, adhering to UI/UX best practices, training large UI models for continuous improvement, and conducting quality assurance and validation. These gaps suggest opportunities for enhancement or the development of new tools that address these specific needs.

## 4. Large user interface (UI) models and their opportunities

Our key contribution in this research (research question three) is the conceptualization of "large user interface models" – LUIMs. In a similar way to LLMs and LVMs, LUIMs work by taking and generating a prediction in the form of user interfaces and usability reports. To the best of our knowledge, previous works focused on generating user interfaces automatically using machine learning models or suggesting the exploitation of large language models (LLMs) to optimize the content of user interfaces. We think this is the first research study to suggest the use of large UI models to facilitate the generation of user interfaces for mobile applications and other interactive systems. Fig. 2. depicts our theoretical model of LUIMs along with its components and the way they interact together to generate and predict favorable UI designs and front-end implementations. For the sake of brevity, Table 4 summarizes the main features of the proposed generative UI model. We suggest LUIMs to significantly capitalize on the existing advantages of large language models, large vision models, and large code models. The synergy between these AI generative models, coupled with UI/UX patterns, scraped UI designs, and user activities and behaviors, promises to produce unimaginable AI designs and experiences.

Table 4. Components of large user interface models and their roles to enable UI generation.

| Component | Input | Key tasks and outputs |
|---|---|---|
| Large UI models (yet to exist) | UI/UX patterns, massive, scraped UI data, user sensory input | Generate new UIs, predict user behavior, predict usability, suggest improvements, improve the user experience (UX) |
| Large language models (e.g., ChatGPT, Bard, Claude, Gemini…) | Text, images, videos, voices | Generate quality content, SEO optimization |
| Large vision models (E.g., Dall-E, Midjourney, Google Imagen, …) | Sketches, themes, layouts | Generate icons and images for UIs, understand sketches and screenshots |
| Large code models (GitHub CoPilot, Amazon CodeWhisperer, …) | Code in different implementation languages | Generate frontend implementations for user interfaces of diverse platforms |

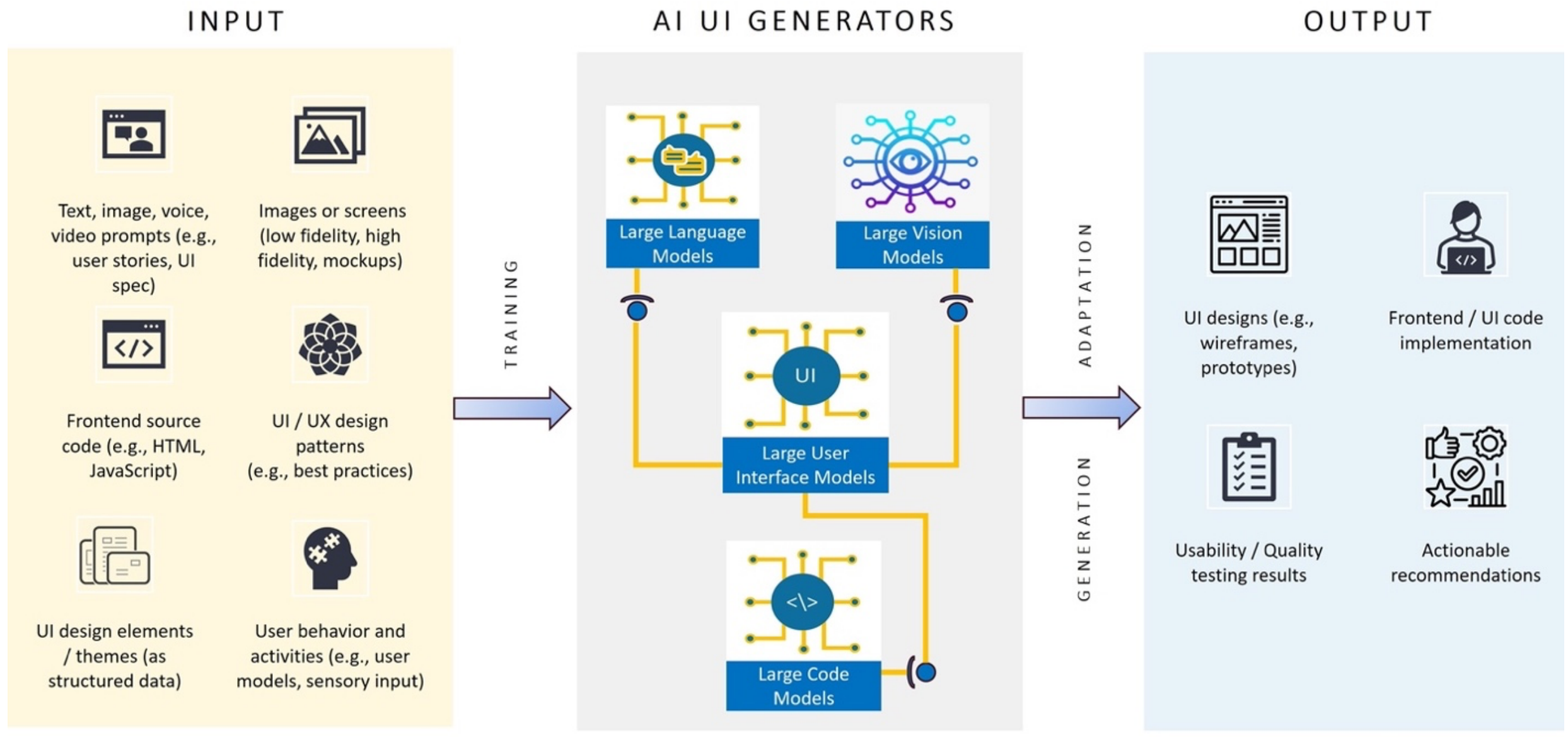

Fig. 2. Our proposed large user interface models (LUIMs), integrating the capabilities of LLMs, LVMs, and LCMs.

We foresee a myriad of opportunities revolutionizing the ways in which we interact with modern technologies, including mobile interfaces. Figure 3 highlights a few examples of how LUIMs could disrupt the software development and digital product design industries and their vast applications in critical domains, such as health and education.

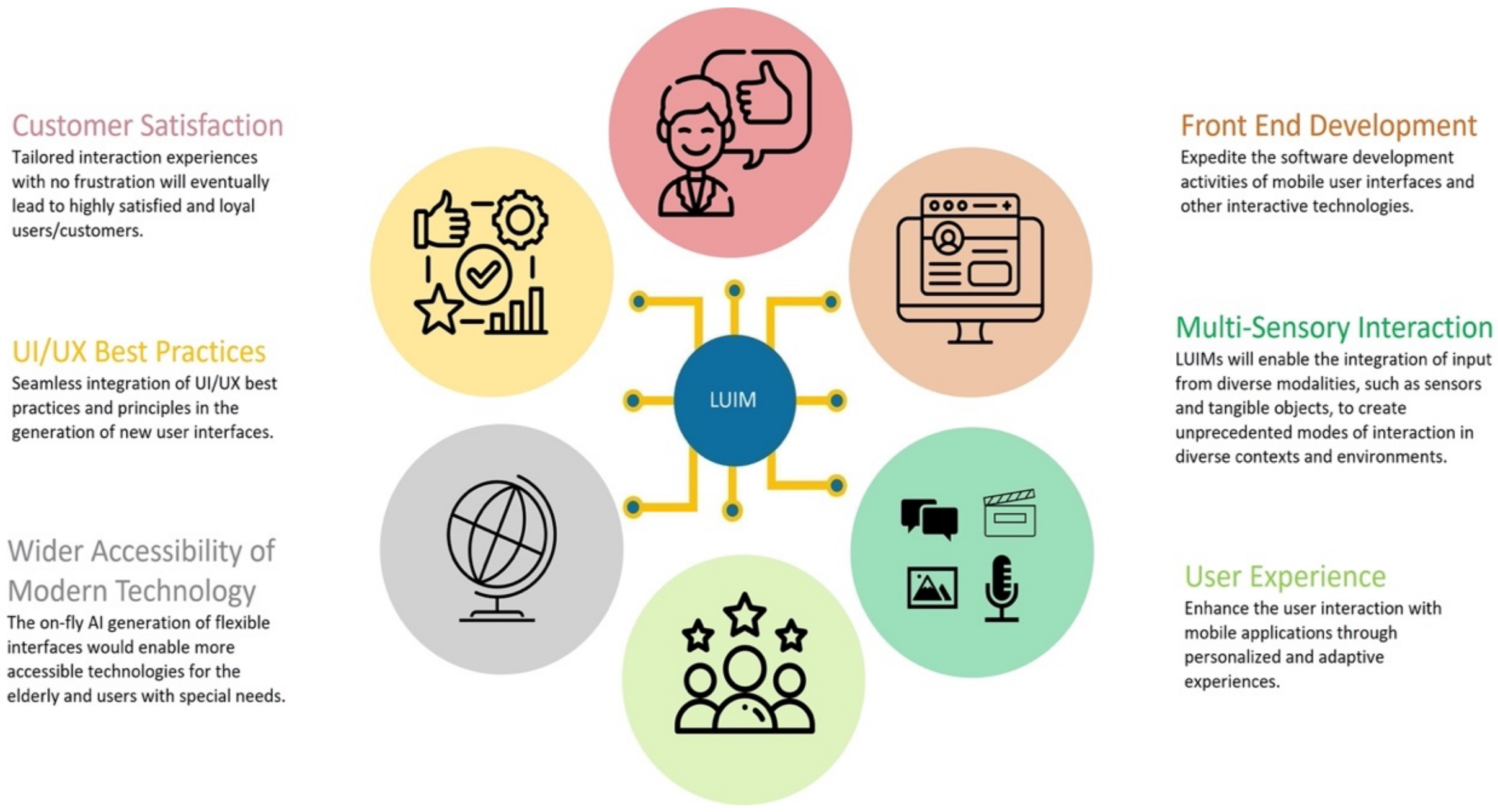

Fig. 3. Opportunities of proposed large user interface models (LUIMs).

## 5. Research challenges concerning large user interface models

Our comprehensive analysis of existing AI-driven mobile usability testing tools and recent studies on generative UIs using deep learning models revealed critical issues and limitations that need to be addressed to enable the creation of robust and satisfying large UI models. We answer the fourth research question by defining the major challenges that LUIMs may encounter.

- Pre-trained lightweight UI generation: one of the key challenges lies in the fact that LVMs require substantial computational resources. Our proposed architecture encompasses language processing, image understanding, and multi-platform coding capabilities, which are all heavy-duty computing tasks. The challenge is to devise a lightweight LUIM that is user-friendly and lightweight.
- UI/UX and multimodal datasets: LUIMs without meaningful UI datasets would not be able to generate acceptable UI designs and predict usability of mobile applications accurately. UI and interaction datasets for diverse mobile technologies need to be collected for the LUIMs.
- Integration of user centric design (UCD) practices: soliciting the intended end users for continuous feedback could prove to be the main challenge of automated large UI models. Software development companies may be inclined to skip the inclusion of users in their development phases. However, synergizing the principles of UCD within the UI generators is a critical factor for the success of mobile interfaces.
- Instilling usability dimensions: it was evident that the latest AI tools focus on the fast generation of visual designs for mobile apps. However, functional elements that define positive experiences such as efficiency, effectiveness, memorability, and learnability, must be captured within the workflow of AI UI generation.
- Comprehension of complex UIs: understanding the semantics of complex mobile layouts and designs will depend on the performance of other models, e.g., large vision models. Moreover, the principles of beautiful UI designs, like symmetry, balance, hierarchy, etc. might be challenging to predict.
- Dynamic personalization of UIs: The LUIMs should empower users to personalize the interfaces as per their dynamic requirements, environments, and mobile platforms while reducing the dependency on coding expertise. Deep learning models may be used to generate mobile interfaces while continuously monitoring the evolving user preferences and contexts.
- Ethical UI/UX design: The LUIMs must be impartial and free from any kind of racial or citizen-based prejudice. When creating user interfaces, it is important to consider user norms, values, and cultures.
- Cross generative AI models integration: the proposed large user interface model is not to be a standalone architecture but is to collaborate with existing LLMs, LVMs, and LCMs. There must be a seamless integration between these models (e.g., through standards, models, etc.) to create business value for mobile development enterprises.

## 6. Conclusion and future directions

This research is the first proposal to create dedicated large UI models (i.e., LUIMs) for the AI generation of graphical user interfaces for mobile apps and other interactive systems. The large user interface models will leverage the powers of large language models (LLMs), large vision models (LVMs), and large code models (LCMs) to generate UIs for mobile conditions and smart technologies. We proposed a LUIM because current user interface design tools lack support for various generative UI/UX design aspects. The existing AI tools do not offer capabilities for training machine learning models. Furthermore, these tools fail to offer guidance on design patterns and best practices regarding usability and user experience. Additionally, the generative UI design tools do not produce output for mobile native applications. Future works will focus on crafting the inner components of the large UI generator and suggest ways to incorporate usability attributes (e.g., efficiency, learnability, effectiveness, satisfaction) into the workflow of the generative UI models.